\documentclass[]{revtex4}
\usepackage{graphicx}

\begin{document}

\title{Spatial coherence singularities and incoherent vortex solitons}

\author{Kristian Motzek}

\affiliation{Institute of Applied Physics, Darmstadt University of
Technology, D-64289 Darmstadt, Germany}

\author{Yuri S. Kivshar}

\affiliation{Nonlinear Physics Center, Research School of Physical
Sciences and Engineering, Australia National University, Canberra
ACT 0200, Australia}

\author{Ming-Feng Shih}

\affiliation{Physics Department, National Taiwan University,
Taipei, 106, Taiwan}

\author{Grover A. Swartzlander, Jr.}

\affiliation{Optical Sciences Center, University of Arizona,
Tucson, Arizona 85721, USA}

\begin{abstract}
We study spatially localized optical vortices created by
self-trapping of partially incoherent light with a phase
dislocation in a biased photorefractive crystal. In a contrast to
the decay of coherent self-trapped vortex beams due to the
azimuthal instability, the incoherent vortices are stabilized when
the spatial incoherence of light exceeds a certain threshold. We
analyze the spatial coherence properties of the incoherent optical
vortices and reveal the existence of ring-like singularities in
the spatial coherence function of a vortex field that can
characterize the stable propagation of vortices through nonlinear
media.
\end{abstract}

\maketitle

\section{Introduction}

Vortices are the fundamental objects in physics, and they can be
found in different types of linear and nonlinear coherent systems.
A typical scalar vortex has the amplitude vanishing at its center
and a well-defined phase associated with the circulation of
momentum around the helix axis~\cite{berry}. In optics, vortices
are associated with phase dislocations (or phase singularities)
carried by optical beams~\cite{soskin}. The last decade has seen a
resurgence of interest in the study of {\em optical
vortices}~\cite{list}, owing in part to readily available
computer-generated holographic techniques for creating phase
singularities in laser beams.

In a self-focusing nonlinear medium, the singular optical beam
undergoes self-focusing, and it becomes self-trapped creating a
stationary ring-like structure with zero intensity at the center
and a phase singularity~\cite{kruglov}. However, such {\em an
optical vortex soliton} is known to be highly unstable in
self-focusing nonlinear media~\cite{book}, it decays by splitting
into several fundamental (no nodes) solitons flying off the
soliton ring~\cite{firth}. This effect has been observed
experimentally in different nonlinear systems, including the
saturable Kerr-like nonlinear media~\cite{tikh}, biased
photorefractive crystals~\cite{exp3}, and quadratic nonlinear
media~\cite{chi2} operating in the self-focusing regime. This
effect is also expected to occur in other physical systems
including the attractive Bose-Einstein condensates~\cite{saito}.

A number of recent theoretical
studies~\cite{cubic,cubic1,cubic2,cubic3,cubic4,cubic5}, including
the rigorous analysis of linear stability of a self-trapped vortex
beam~\cite{buryak,buryak1}, suggest that the stable propagation of
spatial and spatiotemporal vortex-like stationary structures may
become possible in the models with competing nonlinearities in the
presence of large higher-order defocusing nonlinearity; however
such materials are not yet known and no stable {\em coherent}
vortex solitons have been observed in experiment so far.

Recently, stable propagation of spatially localized optical
vortices in a self-focusing biased nonlinear photorefractive
crystal has been observed experimentally in the case when the
vortices are created by partially incoherent light carrying a
phase dislocation~\cite{JengPRL04}. In particular, it was shown,
both experimentally and theoretically, that single- and
double-charge optical vortices can be stabilized in self-focusing
nonlinear media when the spatial incoherence of light exceeds a
certain threshold.

The successful experimental observation of stable self-trapped
vortex beams created by partially incoherent light call for
additional studies of the specific properties of partially
coherent light carrying phase singularities and propagating in a
nonlinear medium. Indeed, if a vortex-carrying beam is partially
incoherent, the phase front topology is not well defined, and
statistics are required to quantify the vortex phase. In the
incoherent limit neither the helical phase nor the characteristic
zero intensity at the vortex center can be observed.

However, several recent studies have shed light on the question
how phase singularities can be unveiled in incoherent light fields
propagating in linear media~\cite{SchoutenOL03,PalaciosPRL04}. In
particular, Palacios {\em et al.}~\cite{PalaciosPRL04} used both
experimental and numerical techniques to explore how a beam
transmitted through a vortex phase mask changes as the transverse
coherence length at the input of the mask varies. Assuming a
quasi-monochromatic, statistically stationary light source and
ignoring temporal coherence effects, they demonstrated that robust
attributes of the vortex remain in the beam, most prominently in
the form of {\em a ring dislocation} in the cross-correlation
function.

The purpose of this paper is twofold. First, we study numerically
the effect of vortex stabilization through the analysis of {\em
the spatial coherence function} of a vortex beam propagating in a
self-focusing nonlinear medium. We reveal the specific features of
the coherent function and demonstrate its importance for the study
of singular beams in nonlinear media. Second, by applying the
modal theory approach, we provide a deeper physical insight into
the effect of the vortex stabilization by partially coherent light
observed in experiment.

The paper is organized as follows. In Sec.~2 we introduce our
numerical model that is based on the coherent density approach and
describes the propagation of partially incoherent light in a
slow-response nonlinear medium such as a biased photorefractive
crystal. Section~3 demonstrates some examples of the stable
partially incoherent vortex solitons, including the experimental
results. In Sec.~4 we introduce the spatial coherence function and
analyze its properties, whereas Sec.~5 is devoted to a simplified
approach based on the truncated modal expansion. Finally, Sec.~6
concludes the paper.

\section{Model and numerical approach}
\label{sec2}

In order to study numerically the propagation of partially
incoherent optical vortices in a biased photorefractive nonlinear
medium, we employ the coherent density
approach~\cite{ChristodoulidesPRL97}. The coherent density
approach is based on the fact that an incoherent light source can
be thought of as a superposition of (infinitely) many coherent
components $E_j$ that are mutually incoherent, having slightly
different propagation directions:
\begin{equation}
\label{eq:10}
 E(\mathbf{r},t)=\sum_j E_j(\mathbf{r})
e^{i\mathbf{k}_{\perp j} \mathbf{r}} e^{i\gamma_j(t)},
\end{equation}
where $\mathbf{k}_{\perp j}= k(\alpha_j\mathbf{e}_x+
\beta_j\mathbf{e}_y)$ is the transverse wave vector of the $j$-th
component, having direction cosines $\alpha_j$ and $\beta_j$,
$\mathbf{r}=x\mathbf{e}_x+y\mathbf{e}_y$, $\gamma_j(t)$ is a
random variable that changes on the time scale of the coherence
time of the light source, and $k=2\pi/\lambda$ is the wavenumber.
The vortex is introduced via a phase mask at the input face
($z=0$) of the medium. To avoid complexities that may arise from
incoherent light sources having abrupt boundaries, we assume the
source has a Gaussian profile
\begin{eqnarray}
  \label{eq:1}
  E_j(\mathbf{r})=\left(\frac{1}{\sqrt{\pi}\theta_0}e^{-(\alpha_j^2+
      \beta_j^2)/\theta_0^2}\right)^{1/2} A(\mathbf{r}),
\end{eqnarray}
where
\begin{equation}
\label{eq:20}
 A(\mathbf{r})=(r/w_0)^2 e^{im\varphi}
e^{-r^2/\sigma^2}
\end{equation}
is the complex vortex profile, $\varphi$ is the angular variable,
and $\theta_0$ is a parameter that controls the beam's coherence,
i.e. less coherence means lager value of $\theta_0$.

Scaling the lengths in the transverse directions to $x_0=1\mu m$
and the length in propagation direction to $z_0=2kx_0^2$, where we
chose $k=2\pi/(230nm)$, the propagating field $E_j(\mathbf{r},z)$
can be described by the nonlinear Schr\"odinger equation:
\begin{eqnarray}
  \label{eq:2}
  i\frac{\partial  E_j(\mathbf{r},z)}{\partial z} +
  \nabla_\perp^2 E_j(\mathbf{r},z) + \eta(\mathbf{r},z)E_j(\mathbf{r},z) =0\, ,
\end{eqnarray}
where $\eta(\mathbf{r},z)$ accounts for the nonlinear refractive
index change in the material. We assume a photorefractive medium
with a saturable nonlinearity having a response time much larger
than the coherence time of the light source. In this case $\eta$
depends on the time-integrated intensity, $I=\sum_j|E_j|^2$, and
it can be written as
\begin{equation}
\label{eq:200}
\eta(\mathbf{r},z)=\frac{I(\mathbf{r},z)}{1+sI(\mathbf{r},z)},
\end{equation}
where $s$ is the saturation parameter. Whereas numerical solutions
of Eq.~(\ref{eq:2}) may be readily computed using the coherence
density approach, later we adopt also the equivalent multi-mode
theory~\cite{ChristodoulidesPRE01} to provide a physical insight
for our findings.

\section{Partially incoherent vortex solitons}
\label{sec3}

The experimental results, first reported in Ref.~\cite{JengPRL04},
were obtained for a vortex beam generated in a self-focusing
biased photorefractive SBN crystal. The rotating diffuser was used
to introduce random-varying phase and amplitude of the input light
beam on the time scales much shorter than the response time of the
crystal. By adjusting the position of the diffuser to near (away
from) the focal point of the lens in front the diffuser, the
degree of the light coherence was increased (decreased). The light
after the rotating diffuser was sent through a computer-generated
hologram to imprint a vortex phase on the light beam. Such a
partially coherent vortex beam was sent into the photorefractive
crystal.

\begin{figure}[htbp]
  \centerline{
    \scalebox{0.4}{
      \includegraphics{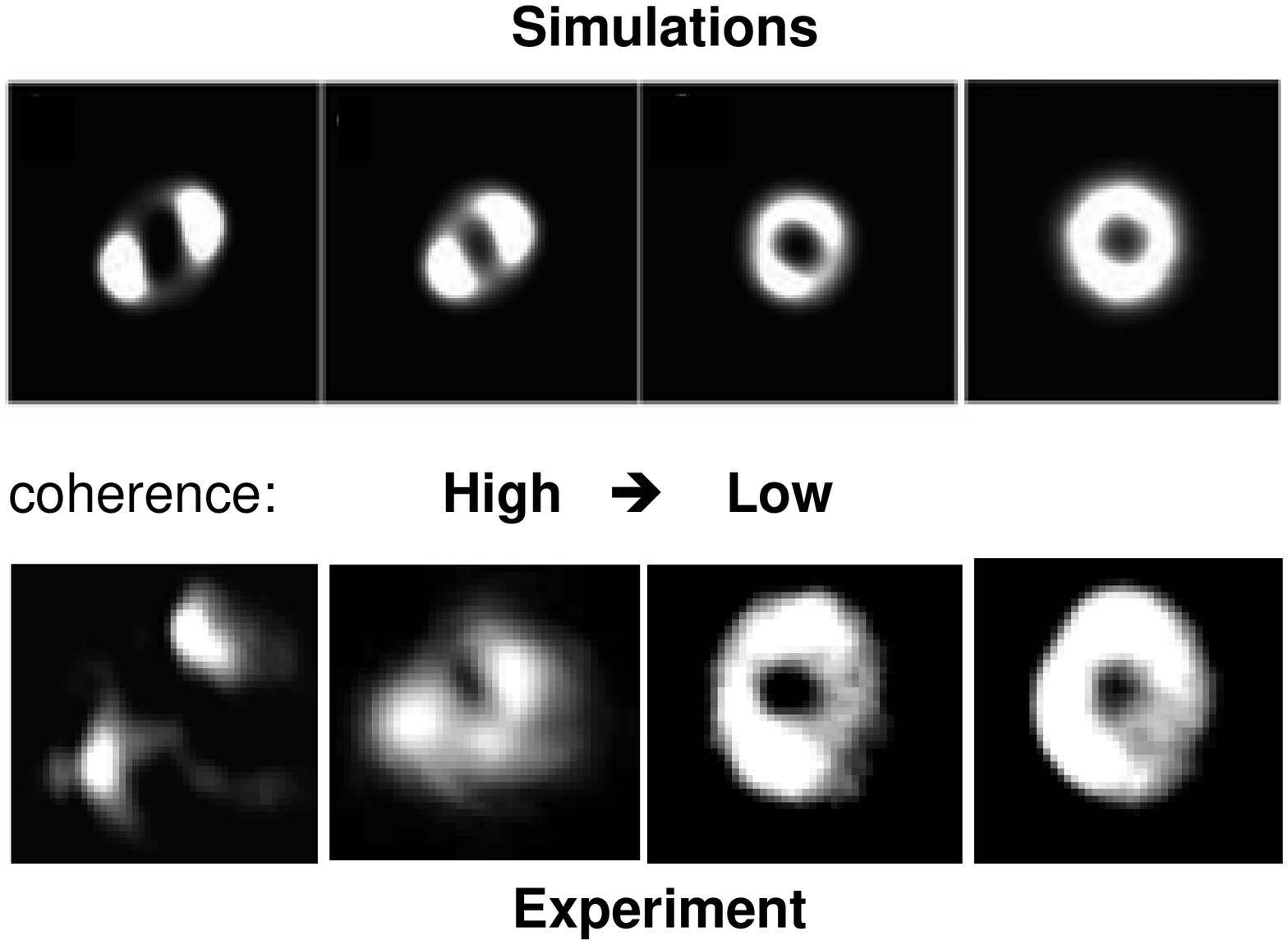}
    }
  }
  \caption{Comparison between numerical (upper row) and
experimental (lower row) results for the vortex stabilization
effect. Numerical results are shown for the vortex after 9mm of
propagation for (from left to right): the coherent case and for
the partially incoherent cases at $\theta_0=0.14$,
$\theta_0=0.29$, $\theta_0=0.38$, respectively. }
  \label{fig0}
\end{figure}

The experimental results are summarized in Fig.~\ref{fig0}(lower
row), and they are compared with the corresponding numerical
results [see Fig.~\ref{fig0}(upper row)] obtained in the framework
of the theoretical model introduced in Sec.~2 above. First, both
numerics and experiment reproduce the well known result that the
coherent single-charge ($m=1$) vortex beam cannot propagate stably
in a self-focusing nonlinear medium (left plots). Indeed, when the
diffuser is removed from the experimental setup and a 2.5 kV
biasing voltage is applied on the photorefractive crystal creating
a Kerr-type self-focusing nonlinear medium, the vortex beam breaks
up into two pieces. This vortex break-up observed in a
self-focusing medium is due to the azimuthal instability, and it
has been observed previously.

When the rotating diffuser is used, the degree of coherence of the
vortex beam varies, and we observe clearly that the vortex beam
can be stabilized by the reduction of the degree of coherence, as
is summarized in Fig.~\ref{fig0}. Above a certain value of the
coherence parameter $\theta_0$, the generated stable partially
incoherent vortex soliton is observed at the output face of the
crystal.

\section{Spatial coherence function}
\label{sec4}

In order to quantify the second-order coherence properties of the
singular beam propagating in a nonlinear medium, we calculate the
mutual coherence function
\begin{equation}
\Gamma(\mathbf{r}_1,\mathbf{r}_2;z) =
\left<E^*(\mathbf{r}_2,z,t)E(\mathbf{r}_1,z,t)\right>,
\end{equation}
where the brackets stand for averaging over the net field
$E(\mathbf{r},z,t) = \sum_{j=1}^N E_j(\mathbf{r},z)
\exp(i\gamma_j(t))$. Again, we assume that for the photorefractive
nonlinearities the random phase factors $\gamma_j(t)$ vary on a
timescale much faster than the response time of the medium. For
the linear propagation, Palacios {\em et al.}~\cite{PalaciosPRL04}
demonstrated that the phase singularities  occur in the
cross-correlation $\Gamma(-\mathbf{r},\mathbf{r})$ of an
incoherent vortex beam, where the origin of the coordinate system
is chosen to coincide with the vortex center.

\begin{figure}[htbp]
  \centerline{
    \scalebox{0.5}{
      \includegraphics{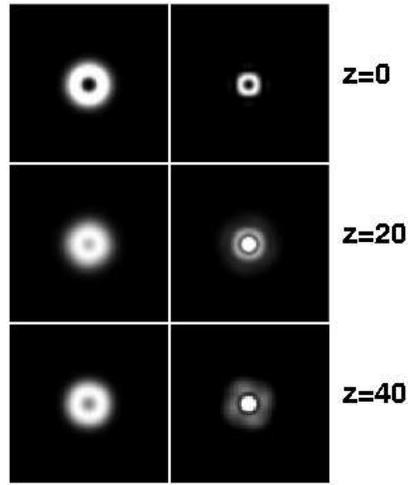}
    }
  }
  \caption{
    Contour plots of the intensity (left column) and the modulus of the cross-correlation
    (right column) of an incoherent vortex with $\theta_0=0.64^\circ$ (strong incoherence).
    Contrary to the case of the linear propagation, there is a local intensity minimum in the
beam's center. The cross-correlation, however, shows the same ring
of phase singularities as predicted in the linear theory.
  }
  \label{fig1}
\end{figure}

\begin{figure}[htbp]
  \centerline{
    \scalebox{0.5}{
      \includegraphics{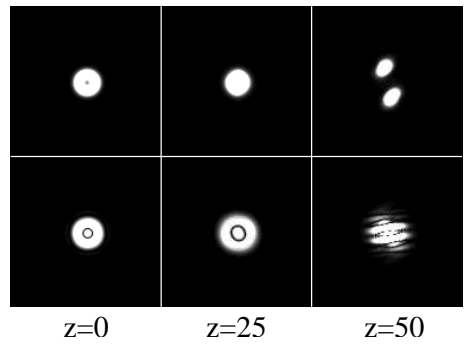}}}
  \caption{
    Contour plots of the intensity (upper row) and the modulus of the cross-correlation
    (lower row) for the breakup of an incoherent vortex at $\theta_0=0.37$ (weak
    incoherence), when the ring is not preserved.
  }
  \label{fig3}
\end{figure}

In Figs.~\ref{fig1}, \ref{fig3}, we show the numerical results for
the stable and unstable nonlinear evolution of an incoherent
vortex and the corresponding evolution of the vortex
cross-correlation function. In these examples, we simulate the
model (\ref{eq:10})-(\ref{eq:200}) with $N=1681$ components, with
the parameters $w_0=1.8$, $\sigma=1.5$ and $s=0.5$. The size of
the numerical simulation domain corresponds to the domain
$35\times35\mu m$.

First, we notice that in the nonlinear case the beam intensity has
a local minimum in the center of the vortex, even after
propagating many diffraction lengths. This is contrary to the case
of the linear propagation where a beam with the same degree of
coherence $\theta_0$ has maximum intensity in the center of the
vortex after only a few diffraction lengths. Also, if we had
chosen to propagate an incoherent ring of light without
topological charge instead of an incoherent vortex, we would also
observe a maximum in the beam's center. Thus we can state that the
coherence function of the vortex manifests itself in the intensity
distribution of the light beam after propagating through a
nonlinear medium. In fact, the intensity profile remains
reminiscent of a vortex, even if the intensity does not quite drop
to zero in the center of the beam.

Analyzing the structure of the beam cross-correlation function, we
clearly observe, similar to the case of the linear
propagation~\cite{PalaciosPRL04}, a ring of phase singularities in
the cross-correlation function $\Gamma(-\mathbf{r},\mathbf{r})$
that is preserved when the vortex is stabilized (see
Fig.~\ref{fig1}) or disintegrates and decays when the vortex
breaks up~(see Fig.~\ref{fig3}).

Thus, as the first result of our numerical studies we state that
the phase singularities in cross-correlation predicted for the
incoherent vortices propagating in linear media also survive the
propagation through a nonlinear medium. This is not self-evident,
considering that in the nonlinear case the separate components
that form an incoherent light beam do interact, contrary to the
linear case. A physically intuitive explanation how this ring of
phase singularities develops under linear propagation is given in
Ref.~[\onlinecite{PalaciosPRL04}]. However,  this issue becomes
more complicated for the propagation in a nonlinear medium.

\begin{figure}[htbp]
  \centerline{
    \scalebox{0.5}{
      \includegraphics{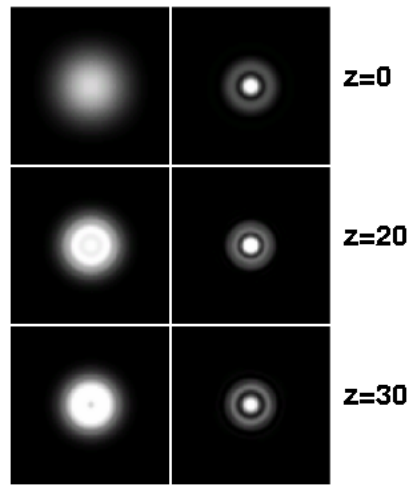}}}
  \caption{
    The intensity (left column) and the cross-correlation (right column) of the
    far field. The effects of the nonlinearity on the intensity distribution
    can be clearly seen, whereas the cross-correlation maintains more or less
    the structure one would expect in the case of linear propagation.
  }
  \label{fig30}
\end{figure}

In addition, in Fig.~\ref{fig30} we show the situation in the far
field. All parameters are identical to those used in
Fig.~\ref{fig1}. In the far field  we observe as well a ring-like
structure of the cross-correlation function
$\Gamma(-\mathbf{f},\mathbf{f})$, where $\mathbf{f}$ stands now
for the spatial coordinates in the far field. The intensity
distribution in the far field can also show a local minimum in the
center of the beam, contrary to what one would obtain if the
vortex is propagating through a linear
medium~\cite{PalaciosPRL04}, and also in contrast to the result we
would obtain if we were propagating a light beam without
topological charge. This emphasizes the importance of the
interaction between the beam coherence function and the
nonlinearity.

\section{Modal theory approach}
\label{sec5}

Although the coherence density approach can be used to simulate
the propagation of partially incoherent light with an arbitrary
accuracy, it is of a little use when it comes to finding an
explanation for the results obtained from the numerical
simulations such as those presented above. A deeper physical
insight can be obtained by using the modal theory of incoherent
solitons~\cite{MitchellPRL97}. According to the modal theory, the
incoherent solitons can be regarded as an incoherent superposition
of guided modes of the waveguide induced by the total light
intensity. Since the incoherent vortices that we are dealing with
induce circularly symmetric waveguides, the guided modes we have
to consider are also circularly symmetric. To explain our
numerical findings, we construct numerically, using a standard
relaxation technique~\cite{muell}, {\em a partially incoherent
vortex soliton} that consists of the circularly symmetric modes
with the topological charges $m=0,\,1$ and $2$:
$E(\mathbf{r})=\sum_{m=0}^2
E_m(\mathbf{r})\exp(im\varphi)\exp(i\gamma_m(t))$. A more precise
modelling of incoherent vortices would require more modes. Here,
we restrict ourselves to {\em three modes} only, assuming that for
a partially incoherent vortex the $m=1$ component should be
dominant and that the next strongest components should be those
with topological charge $m^\prime=m\pm 1$, i.e. $m^\prime=0, \,
2$. Indeed, we find that the main features of incoherent vortex
solitons can be  explained qualitatively using only these three
modes.

\begin{figure}[htbp]
  \centerline{
    \scalebox{0.55}{
      \includegraphics{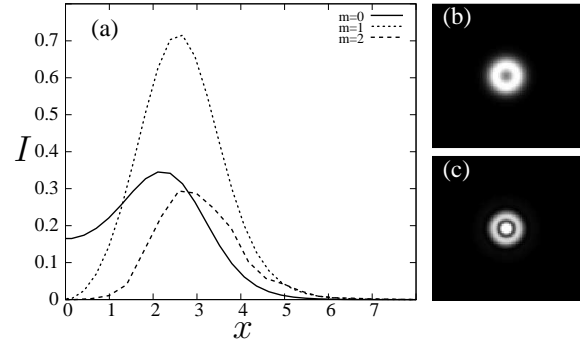}
    }
  }
  \caption{
    A composite soliton calculated by using the three
    modes with the topological charges $m=0,\, 1$ and $3$: (a)
    profiles of the three components, (b) total intensity
    of the vortex soliton, and (c) vortex cross-correlation
    $\Gamma(-\mathbf{r},\mathbf{r})$.
    }
  \label{fig2}
\end{figure}

For this three-mode composite vortex soliton, the relative
intensity of the $m=0$ and $m=2$ modes,  controls the overall beam
coherence, as compared to the $m=1$ main vortex mode. However, in
order to assure that the total topological charge of the beam
\begin{equation}
m_{\rm tot}=\mbox{Im} \{\left<\int E^*(\mathbf{r}\times\nabla
E)\mbox{d}\mathbf{r}\right>\}\mathbf{e}_z/ \int
I\mbox{d}\mathbf{r},
\end{equation}
is equal to one, we have to chose the $m=0$ and $m=2$ components
of equal intensity. In order to check whether this simple approach
yields the results that agree at least qualitatively with the full
numerical model of an incoherent vortex soliton, we calculate the
resulting shape of the vortex components, the total intensity, and
cross-correlation $\Gamma(-\mathbf{r},\mathbf{r})$ shown in
Fig.~\ref{fig2}. Comparing Fig.~\ref{fig1} and Fig.~\ref{fig2}, we
notice the presence of two similar features: (i) the local minimum
of the intensity in the center of the beam, and (ii) the ring-like
structure of the cross-correlation. Hence, these two phenomena can
be explained by considering a simple modal representation of the
incoherent vortex consisting of only three modes with the
topological charges $m=0$, $m=1$, and $m=2$.

First, the local minimum in the center of the beam can be
explained by the fact that the waveguide induced by the $m=1$ and
$m=2$ components affects the $m=0$ mode in such a way, that it
also develops a local intensity minimum in its center, a fact well
known from the vortex-mode vector solitons~\cite{book}. Second,
the ring-like structure of the cross-correlation comes from the
different radial extent of the single components. As is known from
the physics of vortex-mode vector  solitons~\cite{book}, the $m=0$
component has the smallest radial extent, whereas the $m=1$ and
$m=2$ components have larger radii. Hence the cross-correlation
given by $\Gamma(-\mathbf{r},\mathbf{r})=\sum_{m,m^\prime=0}^2
\left<E^*_{m^\prime}(-\mathbf{r})E_m(\mathbf{r})\right>=
\sum_{m=0}^2 E^*_m(-\mathbf{r}) E_m(\mathbf{r})$, is dominated for
small $\mathbf{r}$ by the auto-correlated $m=0$ component, whereas
the $m=1$ component dominates for larger $\mathbf{r}$. For even
larger $\mathbf{r}$, the $m=2$ component can also come into play
which can eventually result in a second ring of auto-correlation.

\section{Conclusions}
\label{sec6}

We have demonstrated stable propagation of optical vortices in a
self-focusing nonlinear medium when the vortices are created by
self-trapped partially incoherent light with a phase singularity
propagating in a slow-response nonlinear medium such as a
photorefractive crystal. The vortex azimuthal instability is found
to be suppressed for the light incoherence above a critical value.
In order to get a deeper physical insight into the effect observed
in both numerics and experiment, we have studied the phase
singularities in the spatial coherence function employed earlier
in the linear optics and demonstrated that they survive the
propagation through nonlinear media when the singular beam creates
an incoherent vortex soliton. Our results emphasize the importance
of the spatial coherence function in the studies of the
propagation of incoherent singular beams. Not only the phase
structure, but also the intensity distribution strongly depends on
the initial form of the coherence function of the light beam as it
enters a nonlinear medium.

\section*{Acknowledgements}

This work was supported by the Australian Research Council and the
Alexander von Humboldt Foundation. Kristian Motzek thanks
Nonlinear Physics Center of the Australian National University for
a warm hospitality.

\newpage

\newpage

\newpage

\newpage

\newpage

\end{document}